\documentclass[12pt]{article}
\usepackage{epsfig}
\usepackage[latin1]{inputenc}
\usepackage{latexsym, amsmath, amssymb, graphicx, amsxtra,euscript,amsfonts,amsthm}
\usepackage{fancyhdr}
\usepackage{amsmath}
\usepackage{amsfonts}
\usepackage{graphicx}
\usepackage{fancyhdr}
\begin{document}
\title{\bf{Higher Order Theories and its Relationship with Noncommutativity}}         
\author{ Oscar S\'anchez-Santos\thanks{oscarsanbuzz@yahoo.com.mx}
\\[0.5cm]
\it Departamento de F\'isica,\\
\it Universidad Aut\'onoma Metropolitana-Iztapalapa\\
\it San Rafael Atlixco 186, C.P. 09340, M\'exico D.F., M\'exico.\\[0.3cm] \\
and \\
Jos\'e David Vergara
\thanks{vergara@nucleares.unam.mx}
\\[0.5cm]
\it Instituto de Ciencias Nucleares,\\
\it Universidad Nacional Aut\'onoma de M\'exico,\\
\it Apartado Postal 70-543, M\'exico 04510 DF, M\'exico
} 
\date{}          
\maketitle
\begin{abstract}
We present a relationship between noncommutativity and higher order
time derivative theories using a method perturbative. We introduce a
generalization of the Chern-Simons Quantum Mechanics for higher
order time derivatives. This model presents noncommutativity in a
natural way when we project to states of low energy. Compared with
the usual model, our system presents noncommutativity without the
necessity of taking the limit of strong field. We quantized the
theory using a Bopp's shift of the noncommutative variables and we
obtain an spectrum without negatives energies. In addition we extend
the model to high order derivatives and noncommutativity with
variable dependent parameter.
\end{abstract}
\section{Introduction}
Theories with higher order time derivatives occur naturally in
several areas of physics \cite{stel,El,Tai}. However a
characteristic of the ordinary Hamiltonian version of these theories
is that this Hamiltonian is linear in the momenta \cite{El} and in
consequence the energy is unbounded from below. However, in most
cases, higher order derivative theories can be treated by
approximation methods \cite{El,Tai}. An essential point in this
construction is the elimination of the high energy degrees of
freedom of the theory and that the symplectic structure is modified
by the procedure. In this work we show that this fact has as
consequence that naturally appears noncommutativity in the system
and so there exist a relationship between noncommutativity and
higher order derivative theories

In order to show the relation between higher order time derivative
theories and noncommutativity we begin by summarizing the theory of
Chern-Simons quantum mechanics and we show how the noncommutativity
arise in the spatial variables.

The theory  of Chern-Simons with derivatives of first order  \cite{Jak}
describes a point particle of mass $m$, confined  to a
quadratic potential and it moves in a plane perpendicular to a
magnetic field, the Lagrangian of the system is given by
\begin{equation}\label{1}
L={m\over2}{\dot x_i}^{2}-{\kappa\over 2}
x_i^{2}+\alpha{\epsilon}_{ij}{x}_{i}{\dot  x}_{j}\mbox{,}
\end{equation}
where, $\epsilon_{ij}$ is the two-dimensional Levy-Civita symbol. In
the limit of zero mass the system is reduced to
\begin{equation}\label{CS1}
L_0=\alpha{\epsilon}_{ij}{x}_{i}{\dot  x}_{j}-{\kappa\over 2}
x^{2}\mbox{.}
\end{equation}
Now, following the Dirac's method of quantization with constraints
\cite{Tei}, we obtain the momenta
\begin{equation}
p_i=-\alpha{\epsilon}_{ij}{x}_{j},
\end{equation}
and the constraints
\begin{equation}
\chi_i=p_i+\alpha{\epsilon}_{ij}{x}_{j}\approx 0.
\end{equation}
The evolution of these constraints do not generate more constraints
then by computing the Poisson brackets we obtain
\begin{equation}
\{\chi_i, \chi_j\}=2{\alpha}{\epsilon}_{ij}.
\end{equation}
We can see that this matrix is invertible then according to the
Dirac formalism that means that we have second-class constraints, in
consequence the symplectic structure is given by the Dirac brackets
\begin{equation}
\{A, B\}_D=\{A, B\}+{1\over {2\alpha}}\{A,
\chi_i\}{\epsilon}_{ij}\{\chi_j, B\}.
\end{equation}
Then, following Dirac procedure we promote this brackets to
commutators in the quantum theory, then the Heisenberg algebra for
the system (\ref{CS1}) is
\begin{equation}
[x_i, p_j]=i\frac{\delta_{ij}}{2}\mathbb{I},\qquad [x_i,
x_j]=-i{{\epsilon}_{ij}\over {2\alpha}},\qquad
[p_i,p_j]=-i{\alpha{\epsilon}_{ij}\over {2}}.
\end{equation}
where we have done $\hbar=1$. In consequence in the limit of zero mass we have noncommutativity in
the spatial variables.

\section{Lagrangian and  Constraints}

Now, the general idea of this paper is to see if it is possible
generalize the above result to the case of theories with high order
time derivatives. First, we will consider an extension of the
Chern-Simons Quantum Mechanics to a second order time derivative
theory \cite{Jer}, with an additional harmonic term, the Lagrangian
chosen has the form
\begin{equation}\label {EC11}
L={m\over2}{\dot x_i}^{2}-{\kappa\over 2} x_i^{2}+\alpha{\epsilon}_{ij}{\dot  x}_{i}{\ddot  x}_{j}\mbox{.}
\end{equation}
In order to study the canonical formalism of this theory we follow the
Ostrogradski procedure in this case the generalized canonical momenta are
defined  by
\begin{equation}\label{EC21}
p_{i}={\partial L\over\partial {\dot x_{i}}}-{d\over dt}\Bigl({\partial L\over\partial {\ddot x_{i}}}\Bigl)\hspace{0.5cm}\mbox{and}\hspace{0.5cm}{\pi}_{i}={\partial L\over\partial {\ddot x_{i}}}\mbox{,}
\end{equation}
For the Lagrangian (\ref {EC11}), one finds
\begin{equation}\label{EC22}
p_{i}=m\dot x_{i}+2\alpha{\epsilon}_{ij}\ddot
x_{j}\hspace{0.5cm}\mbox{and}\hspace{0.5cm}{\pi}_{i}=
-\alpha{\epsilon}_{ij}\dot x_{j}\mbox{.}
\end{equation}
For this theory our phase space is defined by $(x_{i},\dot x_{i},
p_{i},\pi_{i})$, i.e. this theory has, in principle, a higher number
of degrees of freedom. However, due to the equations (\ref{EC22})
the variables of the phase space are not independent, then we have
constraints. These constrains are
\begin{equation}\label{EC23}
\phi_{i}={\pi}_{i}+\alpha{\epsilon}_{ij}\dot x_{j}\mbox{.}
\end{equation}
On the other hand, according to the Ostrogradski formalism the
canonical Hamiltonian, is given by
\begin{equation}\label{EC211}
H_{c}={p_{i}\dot x_{i}\over 2}+{\kappa \over 2}x_{i}^2+
{m\over 2\alpha}\epsilon_{ij}\pi_{i}\dot x_{j}-{\epsilon_{ij}\over 2\alpha}\pi_{i}p_{j}\mbox{.}
\end{equation}
We can see that the first term of this Hamiltonian is linear in the
momenta $p_i$, this show us that the Hamiltonian is unbounded from
below. In the next section we will show how to fix this problem, to
finish this section, we compute the evolution of the constraints
using the total Hamiltonian given by
\begin{equation}\label{EC212}
H={p_{i}\dot x_{i}\over 2}+{\kappa \over 2}x_{i}^2+{m\over 2\alpha}\epsilon_{ij}\pi_{i}\dot x_{j}-
{\epsilon_{ij}\over 2\alpha}\pi_{i}p_{j}+\lambda_{i}\phi_{i}\mbox{.}
\end{equation}
Using this Hamiltonian,  the evolution of the constraints results
\begin{equation}\label{EC213}
\dot\phi_{i}=\{\phi_{i},H\}={\epsilon_{ij}\over
2\alpha}\phi_{j}+2\alpha\epsilon_{ik}\lambda_{k}\approx 0.
\end{equation}
From the above equation we can determine the Lagrange multipliers
\[
\lambda_{i}\approx-{\phi_{i}\over4\alpha^2}\mbox{,}
\]
and in consequence we don't have more constraints. Furthermore,
these constraints are second class, with the Poisson bracket given
by
\begin{equation}\label {EC25}
\{\phi_{i}\mbox{,}\phi_{j}\}=2\alpha\epsilon_{ij}\mbox{,}
\end{equation}
Now, according of Dirac formalism we have to construct the Dirac
brackets, these take the following form
\begin{equation}\label {EC26}
\{A,B\}_{D}=\{A,B\}-\{A,\phi_{i}\}\{\phi_{i},\phi_{j}\}^{-1}\{\phi_{j},B\}.
\end{equation}
In particular, for our theory we have that the matrix
$\{\phi_{i},\phi_{j}\}^{-1}$, is given by
\begin{equation}\label{EC27}
\{\phi_{i},\phi_{j}\}^{-1}=-{\epsilon_{ij}\over2\alpha}\mbox{,}
\end{equation}
Now, by promoting these brackets to commutators we obtain the
following algebra between our operators
\begin{equation}\label{EC216}
[x_{i},p_{j}]=i\delta_{ij}\mathbb{I}\mbox{,}\hspace{0.5cm}[x_{i},x_{j}]=0,\hspace{0.5cm}[\dot
x_{i},\dot x_{j}]=-{i\over2\alpha}\epsilon_{ij}\mbox{,}
\end{equation}
From this algebra we see that the variables associated with the
velocities are noncommutative. This result was obtained without the
necessity of taking any class of limit in counterpart to the first
order theory.  However, the Hamiltonian associated with higher-order
theory still contains problems with the state of minimum energy, in
addition to this, the quantization can not be done directly because
we have a non-canonical algebra. In the next subsection we will show
how to resolve these two problems.
\subsection{Perturbative Approximation and Quantum Spectrum}
In order to obtain a theory without high order time derivatives
in our model and in this way eliminate the states of negative
energy. We will use the perturbative method proposed in \cite{Tai}.
This method will allows us to write the terms with high order
derivatives in terms of first order derivatives. The following
scheme will be used in the next sections, so we review this
procedure. The equations of motion for the Lagrangian (\ref{EC11}) are
\begin{equation}
\ddot x_i=-{\kappa\over m}x_i-{2\alpha\over m}\epsilon_{ij}x_j^{(3)}.
\end{equation}
Now we assume that the contribution of the high order term is weaker
than the other terms in the Lagrangian, consequently we make the
assumption that $\alpha <<1$. Then, the second order time
derivatives can be approached as
\begin{equation}
\ddot x_i\approx-\biggl({\kappa\over m}+{4\alpha^2\kappa^2\over
m^4}\biggl) x_i+\biggl({2\alpha\kappa\over
m^2}+{16\alpha^3\kappa^2\over m^5}\biggl)\epsilon_{ij}\dot
x_j+{\cal{O}}(\alpha^4).
\end{equation}
Higher orders in $\alpha$, are obtained by iterating the equations
of motion. The next step is to built the symplectic form, by using
the brackets (\ref {EC216}) and the constraints (\ref{EC23}) we
obtain
\begin{equation}\label {EC34-1}
\Omega={\omega_{AB}\over 2}dz^{A}\wedge dz^{B}={\delta}_{ij}dp_i\wedge dx_j
+\alpha{\epsilon}_{ij}d\dot x_i\wedge d\dot x_j.
\end{equation}
Now, with our approximations the momenta (\ref{EC22}) are given to
order $\alpha^3$ as
\begin{equation}\label {EC31}
p_{i}=\biggl(m-{4\alpha^2\kappa\over m^2}\biggl)\dot
x_{i}-\biggl({2\alpha\kappa\over m}+{8\alpha^3\kappa^2\over
m^4}\biggl) \epsilon_{ij}x_j+ {\cal O}(\alpha^4),
\hspace{1.0cm}\pi_{i}=-\alpha\epsilon_{ij}\dot x_{j}\mbox{.}
\end{equation}
Introducing the above momenta  (\ref{EC31}) in the symplectic form
(\ref{EC34-1}), we obtain
\begin{equation}\label {EC33}
\Omega=\biggl(m-{4\alpha^2\kappa\over m^2}\biggl)\delta_{ij}d\dot
x_{i}\wedge dx_{j}+ \biggl({2\alpha\kappa\over
m}+{8\alpha^3\kappa^2\over m^4}\biggl)\epsilon_{ij}dx_{i}\wedge
dx_{j}+ \alpha\epsilon_{ij}d\dot x_{i}\wedge d\dot x_{j}+{\cal
O}(\alpha^4)\mbox{.}
\end{equation}
This two-form is the approximation to order $\alpha^3$ to the
symplectic structure. In matrix form $\omega_{AB}$ and its inverse
$\omega^{AB}$, are given by
\begin{equation}\label {EC35}
{\omega_{AB}}=\left (
\begin{array}{rrrr}
0 & {4\alpha\kappa\over m}+{16\alpha^3\kappa^2\over m^4} & -m+{4\alpha^2\kappa\over m^2} & 0 \\
-{4\alpha\kappa\over m}-{16\alpha^3\kappa^2\over m^4} & 0 & 0 & -m+{4\alpha^2\kappa\over m^2} \\
m-{4\alpha^2\kappa\over m^2} & 0 & 0 & 2\alpha \\
0 & m-{4\alpha^2\kappa\over m^2} & -2\alpha & 0
\end{array}
\right)
\mbox{,}
\end{equation}
\begin{equation}\label{invw}
\hspace{0.5cm}{\omega^{AB}}=\left (
\begin{array}{rrrr}
0 & {2\alpha\over m^2}+{32\alpha^3\kappa\over m^5} & {1\over m}+{12\alpha^3\kappa\over m^4} & 0 \\
-{2\alpha\over m^2}-{32\alpha^3\kappa\over m^5} & 0 & 0 & {1\over m}+{12\alpha^3\kappa\over m^4} \\
-{1\over m}-{12\alpha^3\kappa\over m^4} & 0 & 0 & {4\alpha\kappa\over m^3}+{80\alpha^3\kappa^2\over m^6} \\
0 & -{1\over m}-{12\alpha^3\kappa\over m^4} & -{4\alpha\kappa\over m^3}-{80\alpha^3\kappa^2\over m^6} & 0
\end{array}
\right)
\mbox{.}
\end{equation}
By using the matrix (\ref{invw})  we read the basic new brackets,
$(\omega^{AB})_{ij}=\{z_{i},z_{j}\}_D$ (where $z_i=\{x_1,x_2,\dot x_1,\dot x_2\}$),
explicitly these parenthesis are given by
\begin{align}\label {EC37111}
\{x_{i},x_{j}\}_D=&\biggl({2\alpha\over m^2}+{32\alpha^3\kappa\over m^5}\biggl)\epsilon_{ij}\mbox{,} \ \ \
\{\dot x_{i},\dot x_{j}\}_D=\biggl({4\alpha\kappa\over m^3}+{80\alpha^3\kappa^2\over m^6}\biggl)
\epsilon_{ij}\mbox{,}\\
&\{ x_{i},\dot x_{j}\}_D={1\over m}\biggl(1+{12\alpha^2\kappa\over
m^3}\biggl)\delta_{ij}\mbox{.} \label{EC37.1}
\end{align}
To avoid the additional extra constant factor in (\ref{EC37.1}), we define
\begin{equation}\label{rhoi}
 \rho_i=\bigl(1-{12\alpha^2\kappa\over m^3}\bigl){m\dot x_i},
\end{equation}
in consequence the basic parenthesis in this case are
\begin{equation}\label{EC37}
\{x_{i},x_{j}\}_D=\biggl({2\alpha\over m^2}+{32\alpha^3\kappa\over m^5}\biggl)\epsilon_{ij}\mbox{,}
\hspace{0.5cm}\{\rho_{i},\rho_{j}\}_D=\biggl({4\alpha\kappa\over m}-{16\alpha^3\kappa^2\over m^4}\biggl)
\epsilon_{ij}\mbox{,}
\end{equation}
\[
\{ x_{i},\rho_{j}\}_D=\delta_{ij}\mbox{.}
\]
On the other hand, if we introduce the momenta (\ref{EC31}) in the
Hamiltonian (\ref{EC211}) and the definition of $\rho_i$, we obtain
the Hamiltonian in terms of the new variables. So, to third order in
$\alpha$ we get
\begin{equation}\label{EC310}
H={1\over 2m}\biggl(1+{16\alpha^2\kappa\over m^3}\biggl)\rho_{i}^2+{\kappa\over 2}x_{i}^{2}+
\biggl({2\alpha\kappa\over m^2}+{32\alpha^3\kappa^2\over m^5}\biggl)
\epsilon_{ij}x_{i}\rho_{j}+{\cal O}(\alpha^4)\mbox{.}
\end{equation}
In this way, directly from the high order theory we get a
noncommutative theory with Dirac brackets in the reduced phase space
given by (\ref{EC37}) and Hamiltonian (\ref{EC310}). The interesting
feature of the high order theory (\ref{EC11}) is that contains the
noncommutativity without taking any limit in the kinetical term in
contrast with the first order Chern-Simons quantum mechanics of
(\ref{1}).

Now, the more simple way to quantize this noncommutative theory
is to map noncommutative phase space to the ordinary phase space \cite{Sma, Jian},
which satisfy the following commutation relations
\begin{equation}\label {EC311}
\{\bar x_{i},\bar x_{j}\}=\{{\bar\rho}_{i},{\bar\rho}_{j}\}=
0\mbox{,}\hspace{0.5cm}\{\bar x_{i},{\bar\rho}_{j}\}=\delta_{ij}\mbox{.}
\end{equation}
The mapping that relates the new variables to the old variables is given by
\begin{equation}\label {EC312}
 x_{i}=A_{ij}\bar x_{j}+B_{ij}{\bar\rho}_{j}\mbox{,}\hspace{0.5cm}
 \rho_{i}=C_{ij}\bar x_{j}+D_{ij}{\bar\rho}_{j}\mbox{,}
\end{equation}
In the which $\bf A$, $\bf B$, $\bf C$ and $\bf D$ are $\bf{2\times 2}$
transformation matrices. Following the procedure proposed in \cite{Kang},
one can easily get the conditions that the
transformation matrices should satisfy, these are
\begin{equation}\label {EC314}
\mbox{{A}}_{ik}\mbox{{B}}_{jk}-\mbox{{B}}_{ik}\mbox{{A}}_{jk}=
\biggl({2\alpha\over m^2}+{32\alpha^3\kappa\over m^5}\biggl)\epsilon_{ij}\mbox{,}\hspace{0.5cm}
\mbox{{C}}_{ik}\mbox{{D}}_{jk}-\mbox{{D}}_{ik}\mbox{{C}}_{jk}=
\biggl({4\alpha\kappa\over m}-{16\alpha^3\kappa^2\over m^4}\biggl)\epsilon_{ij}\mbox{,}
\end{equation}
\[
\mbox{{A}}_{ik}\mbox{{D}}_{jk}-\mbox{{B}}_{ik}\mbox{{C}}_{jk}=\delta_{ij}\mbox{.}
\]
If we choose for $\bf A$ and $\bf D$ diagonal matrices so that
$\mbox{{A}}_{ij}=a\delta_{ij}$, $\mbox{{D}}_{ij}=b\delta_{ij}$, and
in addition we select for  {\bf{B}} and {\bf{C}} antisymmetric
matrices, then we get
\begin{equation}\label {EC318}
\mbox{{B}}_{ij}=-{1\over a}\biggl({\alpha\over
m^2}+{16\alpha^3\kappa\over m^5}\biggl)\epsilon_{ij}\mbox{,}
\hspace{0.5cm}\mbox{{C}}_{ij}={1\over b}\biggl({2\alpha\kappa\over
m}-{8\alpha^3\kappa^2\over m^4}\biggl)\epsilon_{ij}
\mbox{,}\hspace{0.5cm}\mbox{{B}}_{ik}\mbox{{C}}_{jk}=(ab-1)
{\delta_{ij}}\mbox{,}
\end{equation}
resolving these set of equations for $b$ up to quadratic order in
$\alpha$, we obtain
\begin{equation}\label {EC320}
b\approx {1\over a}-{2\alpha^2\kappa\over a m^3 } +{\cal
O}(\alpha^4)\mbox{.}
\end{equation}
Therefore, the transformations take the following form
\begin{equation}\label {EC321}
x_{i}=a{\bar x}_{i}-{1\over a}\biggl({\alpha\over m^2}+
{16\alpha^3\kappa\over m^5}\biggl)\epsilon_{ij}{\bar \rho}_{j}+\dots
\mbox{,}\ \ \rho_{i}={1\over a}\left( 1 -{2\alpha^2\kappa\over a m^3
} \right)\bar{\rho}_i +a\biggl({2\alpha\kappa\over
m}-{4\alpha^3\kappa^2\over m^4}\biggl)\epsilon_{ij}{\bar x}_{j}+...
\end{equation}

These transformations (\ref {EC311}) will allow us to quantized our
theory. Introducing the transformations (\ref {EC321}) in the
Hamiltonian (\ref {EC310}), this takes the form
\begin{equation}\label {EC327}
H={\mbox{A}(\alpha,\kappa, m)}{{{\bar\rho}}_{i}}^{2}+\mbox{B}
(\alpha,\kappa, m){{\bar x}_{i}}^{2}+\mbox{C}(\alpha,\kappa, m)
\epsilon_{ij}{\bar x}_{i}{\bar\rho}_{j}\mbox{,}
\end{equation}
where, the constants parameters $(\mbox{A},\mbox{B},\mbox{C})$, up
to third order in $\alpha$, are
\[
\mbox{A}(\alpha,\kappa, m)={1\over 2ma^2}\biggl({1}+
{9\alpha^2\kappa\over m^4}+{\cal
O}(\alpha^4)\biggl),\hspace{0.2cm}\mbox{B}(\alpha,\kappa, m)
=a^2\biggl({\kappa \over 2}-{2\alpha^2\kappa^2\over m^3}+{\cal
O}(\alpha^4)\biggl),
\]
\begin{equation}\label {EC328}
\mbox{C}(\alpha,\kappa, m)=-{\alpha\kappa\over
m^2}-{8\alpha^3\kappa^2\over m^5}+{\cal O}(\alpha^4).
\end{equation}
We recognize (\ref {EC327}) as the Hamiltonian for the commutative,
isotropic 2-dimensional harmonic oscillator, with a coupling term
proportional to the $L_z$ angular momentum. To quantize the theory
we use the coordinate representation $| \bar x_i\rangle$. In this
case the momenta $\bar\rho_{i}$ are promoted directly to operators.
In consequence the Hamiltonian (\ref {EC327}) takes the form
\begin{equation}\label {EC333}
{\hat H}=-A{\partial^{2}\over\partial {\bar x}_{i}^{2}}+B{{\bar x}_{i}}^{2}-
iC\epsilon_{ij}{\bar x}_{i}{\partial\over\partial{\bar x}_{j}}\mbox{,}
\end{equation}
where we have choose the normal ordering. To solve the eigenvalue
problem we write this operator in polar coordinates and it takes the
form
\begin{equation}\label {EC336}
\Biggl[A\Biggl({\partial^{2}\over\partial {r}^{2}}+{1\over r}{\partial\over\partial r}-
{{\hat L}^{2}\over r^{2}}\Biggl)-B{r}^{2}-C{\hat L}\Biggl]\psi(r,\theta)=E\psi(r,\theta)\mbox{.}
\end{equation}
It is convenient to introduce the following redefinition, given in
\cite{Sma},
\begin{equation}\label {EC337}
z=\sqrt{B\over A}r^{2}\mbox{.}
\end{equation}
Using this redefinition the action of the angular momentum operator
results
\begin{equation}\label {EC338}
{\hat L}\psi(r,\theta)=l\psi(r,\theta)=lZ(z)\phi(\theta),
\end{equation}
where $l$ takes the values $0,\pm 1, \pm 2, ...,$, therefore, the
resulting equation for $Z(z)$ is
\begin{equation}\label {EC342}
zZ^{''}(z)+(1-z)Z^{'}(z)+\Bigl[{\mathcal{E}}-{l^{2}\over
4z}\Bigl]Z(z)=0,
\end{equation}
with $\mathcal{E}$ given by
\begin{equation}\label {EC343}
{\mathcal{E}}={1\over4\sqrt{AB}}[E-l C]-{1\over2}.
\end{equation}
The general solution for wave equation is given in terms of the
generalized Laguerre polynomials
\begin{equation}\label {EC344}
\psi_{n_{r},l}(z,\theta)=Nz^{|l|/2}L_{n_{r}}^{|l|}(z)
\exp\Bigl({-{z\over2}+il\theta}\Bigl),
\end{equation}
with
\begin{equation}\label {EC345}
L_{n}^{r}(z)=z^{-r}\exp(z){d^{n}\over dz^{n}}\Bigl(z^{n+r}\exp(-z)\Bigl)\mbox{.}
\end{equation}
Here, $N$ is the proper normalization  constant and $n_r$ is the
radial quantum number. Therefore, in general the quantum spectrum is
given by
\begin{equation}\label {EC346}
E_{n_{r},l}=2\sqrt{AB}(2n_{r}+|l|+1)+l C.
\end{equation}
with quantum numbers taking values $n_r=0,1,2,...$, $l=0,\pm 1,\pm
2,...$. Note that the spectrum only  depends of the constant $a$,
the mass of the system and the parameters $\kappa$, $\alpha$, so
that the spectrum of the system is uniquely determined. Also, this
spectrum has a well-defined minimum energy state. In order to make
clear this fact, we define the following positive numbers
$(n_+,n_-)$\cite{Sma}, which are determined as follows
\begin{equation}
n_r=n_- +{l-|l|\over 2},\hspace{0.5cm}l=n_+-n_-.
\end{equation}
Introducing these quantum numbers in the energy (\ref{EC346}), we obtain
\begin{equation}
E_{n_+,n_-}={\sqrt{\kappa\over m}}\biggl[1+{5\alpha^2\kappa\over
2m^3}+
{\cal{O}}(\alpha^{4})\biggl](n_++n_-+1)+\biggl[{\alpha\kappa\over
m^2}+{8\alpha^3\kappa^2\over m^5}+
{\cal{O}}(\alpha^{5})\biggl](n_+-n_-).
\end{equation}
Therefore, for minimum energy state we get
\begin{equation}
E_{0,0}={\sqrt{\kappa\over m}}\biggl[1+{5\alpha^2\kappa\over 2m^3}+
{\cal{O}}(\alpha^{4})\biggl],
\end{equation}
this energy is positive definite. We can also see that in the limit
$\alpha \to 0$ we recover the usual case of two harmonic
oscillators.
\section{Model of Chern-Simons with Higher Derivatives}
In the previous section was shown that noncommutativity and a high
order derivative theory are closely related, we show this through
the Chern-Simons quantum mechanics of second order. In this section
we will introduce an additional extension of this model, we will
consider now a model with $n$-th order derivatives. With this in
mind the Lagrangian of the model is given by
\begin{equation}\label{EC7.1}
L={m\over 2}\dot x_i^2+{\kappa\over 2}x_i^2+\alpha\epsilon_{ij}x_i^{(n-1)}x_j^{(n)},
\hspace{0.5cm}i=1,2.
\end{equation}
Here $\alpha$  is a constant parameter that measure the high order character of the theory.
According to the Ostrogradski formalism the momenta are defined as
\begin{equation}\label{EC7.2}
p_{mi}=\sum_{k=m}^{n}\biggl(-{d\over dt}\biggl)^{k-m}{\partial L\over\partial x_i^{(k)}}.
\end{equation}
In particular for the case of $m = n$ we have
\begin{equation}\label{EC7.3}
p_{ni}=-\alpha\epsilon_{ij}x_j^{(n-1)}.
\end{equation}
Following Dirac method, the above relation is a constraint, since by Ostrogradski the $n-1$-th derivative
is part of the configuration space, and the full phase space of the theory is
given by ${\{x_i,p_{1i},\dot x_i,p_{2i},\ddot x_i,p_{3i},...,x_i^{(n-1)},p_{ni}\}}$.
 Consequently we get the constraints
\begin{equation}\label{EC7.4}
\phi_i=p_{ni}+\alpha\epsilon_{ij}x_j^{(n-1)}\approx 0.
\end{equation}
We observe that these constraints tell us that the $n$-th momenta
are not independent of each other. Moreover the Poisson brackets
between these constraints are given by
\begin{equation}\label{EC7.5}
\{\phi_i,\phi_j\}=2\alpha\epsilon_{ij}.
\end{equation}
This matrix is invertible and in consequence we are dealing with second class constraints. The
corresponding Dirac brackets are
\begin{equation}\label{EC7.7}
\{A,B\}_D=\{A,B\}+{1\over 2\alpha}\{A,\phi_i\}\epsilon_{ij}\{\phi_j,B\}.
\end{equation}
What follows now is to identify the phase space variables and obtain
the algebra. Making this identification we have the following
non-zero brackets
\begin{equation}\label{EC7.8}
\{x_i,p_{1j}\}_D=\delta_{ij},\dots,
\{x_i^{(n-2)},p_{n-1j}\}_D=\delta_{ij},\hspace{0.5cm}
\{x_i^{(n-1)},x_j^{(n-1)}\}_D=-{\epsilon_{ij}\over 2\alpha}.
\end{equation}
At this point we can conclude by brief inspection that the results
for $n = 2$, corresponds to the previous section. Now, the next step
is to build the symplectic structure, using the Dirac brackets
(\ref{EC7.8}) and applying the second class constraints to obtain a
non degenerate form, the process results in
\begin{equation}\label{EC7.10}
\Omega=\sum_{m=2}^{n-1}dp_{m-1i}\wedge dx_i^{(m-2)}+
\alpha\epsilon_{ij}dx_i^{(n-1)}\wedge dx_j^{(n-1)}.
\end{equation}
To avoid the problems of a Hamiltonian not bounded from below we
apply the perturbative method of \cite{Tai}. The equations of
motion, for this system are
\[
\ddot x_i=-{\kappa\over
m}x_i+(-1)^{n-1}2\alpha\epsilon_{ij}x_j^{(2n-1)}.
\]
To order $\alpha$, we rewrite these equations as
\begin{equation}\label{EC7.11}
\ddot x_i\approx-{\kappa\over m}x_i+{2\alpha\kappa^{n-1}\over m^n}\epsilon_{ij}\dot x_j+{\cal O}(\alpha^2),
\end{equation}
to obtain  higher orders in $\alpha$ we need to iterate the
equations of motion. In general for the high order derivatives, we
get
\[
x_i^{(2k)}\approx\biggl(-{\kappa\over
m}\biggl)^kx_i+(-1)^{k+1}{2k\alpha\kappa^{n+k-2}\over m^{n+k-1}}
\epsilon_{ij}\dot x_j+{\cal O}(\alpha^2),
\]
\begin{equation}\label{EC7.12}
x_i^{(2k+1)}
\approx\biggl(-{\kappa\over m}\biggl)^{k}\dot x_i+(-1)^{k}{2k\alpha\kappa^{n+k-1}\over m^{n+k}}\epsilon_{ij}x_j+{\cal O}(\alpha^2),\hspace{0.5cm}\mbox{with}\hspace{0.5cm}k=1,2,3,....
\end{equation}
On the other hand, the momenta are given by
\[
p_{1i}=m\dot x_i+(-1)^{n}2\alpha\epsilon_{ij}\alpha x_j^{(2n-1)},
\]
\begin{equation}\label{EC7.13}
p_{mi}=(-1)^{n-m-1}2\alpha\epsilon_{ij}\alpha
x_j^{(2n-m-1)},\hspace{0.3cm}\mbox{for}\hspace{0.3cm}m=2,3,4,...,n-1,
\end{equation}
\[
p_{ni}=-\alpha\epsilon_{ij} x_j^{(n-1)},
\]
where $n$ is the order of theory.  As we can see the momenta are
proportional to the time derivatives, and using the approximations
(\ref{EC7.12}), we can replace these derivatives, either by the
positions or by the first time derivative. In this way the
symplectic structure (\ref{EC7.10}), is reduced to
\begin{equation}\label{EC7.14}
\Omega=m\delta_{ij}d\dot x_i\wedge dx_j+{n\alpha\kappa^{n-1}\over m^{n-1}}\epsilon_{ij}
dx_i\wedge dx_j+{(n-1)\alpha\kappa^{n-2}\over m^{n-2}}\epsilon_{ij}d\dot x_i\wedge d\dot x_j.
\end{equation}
Defining $\rho_i=m\dot x_i$ in the same way that in the case of the theory of order two, the symplectic two-form is reduced to
\begin{equation}\label{EC7.15}
\Omega=\delta_{ij}d\rho_i\wedge dx_j+{n\alpha\kappa^{n-1}\over m^{n-1}}
\epsilon_{ij}dx_i\wedge dx_j+{(n-1)\alpha\kappa^{n-2}\over m^{n}}
\epsilon_{ij}d\rho_i\wedge d\rho_j+{\cal O}(\alpha^2).
\end{equation}
Using this two-form we read the Dirac brackets of the reduced theory
\begin{equation}\label{EC7.16}
\{x_i,x_j\}_D={2(n-1)\alpha\kappa^{n-2}\over m^{n}}\epsilon_{ij},
\hspace{0.2cm}\{\rho_i,\rho_j\}_D
={2n\alpha\kappa^{n-1}\over m^{n-1}}
\epsilon_{ij},\hspace{0.2cm}\{x_i,\rho_j\}_D=\delta_{ij},
\end{equation}
these brackets are valid to first order in $\alpha$. The resulting Hamiltonian to first order in $\alpha$,
is given by
\begin{equation}\label{EC7.16.1}
H={\rho_i^2\over 2m}+{\kappa\over 2}x_i^2+{2(n-1)
\alpha\kappa^{n-1}\over m^{n}}\epsilon_{ij}x_i\rho_j+
{\cal O}(\alpha^2),
\end{equation}
As an example we consider the model with third-order derivatives,
which has the following Lagrangian
\begin{equation}\label{EC7.17}
L={m\over 2}\dot x_i^2+{\kappa\over 2}x_i^2+\alpha \epsilon_{ij}x_i^{(2)} x_j^{(3)}.
\end{equation}
The momenta associated with this theory are
\begin{equation}\label{EC7.18}
p_{1i}={\partial L\over\partial x_i^{(1)}}-{d\over dt}
\biggl({\partial L\over\partial x_i^{(2)}}\biggl)+{d^2\over dt^2}
\biggl({\partial L\over\partial x_i^{(3)}}\biggl)=mx_i^{(1)}-
2\alpha\epsilon_{ij}
x_j^{(4)}.
\end{equation}
\begin{equation}\label{EC7.19}
p_{2i}={\partial L\over\partial x_i^{(2)}}-{d\over dt}
\biggl({\partial L\over\partial x_i^{(3)}}\biggl)=2\alpha\epsilon_{ij}
x_j^{(3)}.
\end{equation}
\begin{equation}\label{EC7.20}
p_{3i}={\partial L\over\partial x_i^{(3)}}=-\alpha\epsilon_{ij}x_j^{(2)}.
\end{equation}
As previously anticipated, the momenta associated with the
derivative of highest order, define a constraint in the theory, that
results
\begin{equation}\label{EC7.21}
\phi_i=p_{3i}+\alpha\epsilon_{ij}x_j^{(2)}.
\end{equation}
The resulting Dirac brackets are given by
\begin{equation}\label{EC7.27}
\{x_i,p_{1j}\}_D=\{\dot x_i,p_{2j}\}_D=\delta_{ij}\hspace{0.2cm}
\{\ddot x_i,\ddot x_j\}_D=-{\epsilon_{ij}\over 2\alpha}.
\end{equation}
In consequence, we obtain for the symplectic structure the following expression
\begin{equation}\label{EC7.28}
\Omega=\delta_{ij}dp_{1i}\wedge dx_{j}+\delta_{ij}
dp_{2i}\wedge \dot x_i+\alpha\epsilon_{ij}d\ddot x_i\wedge \ddot x_j.
\end{equation}
On the other hand, to first-order in $\alpha$ the momenta are
reduced to
\[
p_{1i}\approx m\dot x_i-{2\alpha\kappa^2\over m^2}\epsilon_{ij}x_j+
{\cal O}(\alpha^2),
\]
\[
p_{2i}\approx -{2\alpha\kappa\over m}\epsilon_{ij}\dot x_j+{\cal O}(\alpha^2),
\]
\begin{equation}\label{EC7.29}
p_{3i}\approx {\alpha\kappa\over m}\epsilon_{ij}x_j+{\cal O}(\alpha^2).
\end{equation}
With the symplectic two-form and these approximations we have the
following set of Dirac brackets to first order in $\alpha$
\begin{equation}\label{EC7.32}
\{x_i,x_j\}_D={2\alpha\kappa\over m^3}\epsilon_{ij},
\hspace{0.3cm}\{\rho_i,\rho_j\}_D={6\alpha\kappa^2\over m^2}
\epsilon_{ij},\hspace{0.3cm}\{x_i,\rho_j\}_D=\delta_{ij},
\end{equation}
Finally, the reduced Hamiltonian is given by
\begin{equation}\label{EC7.16.111}
H={\rho_i^2\over 2m}+{\kappa\over 2}x_i^2+{4\alpha\kappa^{2}\over m^{3}}
\epsilon_{ij}x_i\rho_j+{\cal O}(\alpha^2).
\end{equation}
So, in this case we get again a noncommutative theory, with the
Dirac's brackets promoted to commutators and a Hamiltonian
equivalent to (\ref{EC310}) and the quantization of this system can
be done following a similar sequence of steps as those used in Sec.
2.1.

\section{Theory with Noncommutative Local Parameter}
So far, starting from a high order time derivative theory we have
obtained a noncommutative theory with constant noncommutative
parameter. The idea of this section is to show that it is possible
to generalize this result to the case of a position dependent
noncommutative parameter. We begin by consider a Lagrangian similar
to the introduced in Ref. \cite{Gomes2}, given by
\begin{equation}\label {EC347}
L={m_r\dot r^2\over 2}-V(r)+{m\over 2}{\dot x_i}^2-{k\over 2}{
x_i}^2 +{\theta f(r)\over 2}{\epsilon}_{ij}{\dot x_i}{\ddot x_j}.
\end{equation}
Noting now that in counterpart of previous example with constant
parameter, in this new theory we aggregate the dynamics in the
variable $r$ and we have done $\alpha=\theta f(r)/2$.

The momenta associate to this Hamiltonian are given by
\begin{equation}\label {EC348}
p_r={m_r\dot r},\hspace{0.5cm}p_i={m}{\dot x_i}+
{\theta f(r)}{\epsilon}_{ij}{\ddot x_j}+{\theta f^{'}(r)\over 2m_r}{p_r}
{\epsilon}_{ij}{\dot x_j},\hspace{0.5cm}\pi_i=-{\theta f(r)\over 2}
{\epsilon}_{ij}{\dot x_j}.
\end{equation}
From the above expressions we observe that we have a constraint,
that results
\begin{equation}\label {EC349}
\chi_i=\pi_i+{\theta f(r)\over 2}{\epsilon}_{ij}{\dot x_j}.
\end{equation}
From the evolution of this constraint we found the associated
Lagrange multiplier, given by
\begin{equation}
\lambda_i\approx{f^{'}(r)p_r\over 2m\theta
f(r)^2}\epsilon_{ij}\pi_j-{m\pi_i\over \theta^2
f(r)^2}-{f^{'}(r)p_r\rho_i\over 4mm_r\theta
f(r)}+{f^{'}(r)^2\pi_i\rho_i^2\over
4m^2m_rf(r)^2}-{\epsilon_{ij}\rho_j\over 2\theta f(r)}.
\end{equation}
Note that the Lagrange multiplier is well defined only if the
function $f(r)$ does not vanish in the interval of definition of the
$r$ variable. This is more clearly seen, from the Poisson bracket of
the constraints, where we get
\begin{equation}\label {EC350}
\{\chi_i,\chi_j\}=\theta f(r){\epsilon}_{ij}.
\end{equation}
So, for nonvanishing $f(r)$ we obtain second class constraints.
Therefore, we can proceed to construct the Dirac brackets
\begin{equation}\label {EC351}
\{A,B\}_D=\{A,B\}+\{A,\chi_i\}{{\epsilon}_{ij}\over \theta f(r)}\{\chi_j,B\}.
\end{equation}
Our phase space is formed by $z^{A}=\{r,x_i,\dot
x_i,p_r,p_i,{\pi}_{i}\}$. Working with the constraints the algebra
of the reduced phase space is
\begin{equation}\label {EC356}
\{r,p_r\}_D=1,\hspace{0.5cm}\{x_i,p_j\}_D=\delta_{ij},
\hspace{0.5cm}\{\dot x_i,\dot x_j\}_D=-{\epsilon_{ij}\over \theta f(r)},
\hspace{0.5cm}\{p_r,\dot x_i\}_D={f^{'}(r)\over 2f(r)}\dot x_i,
\end{equation}
By using the above algebra we obtain the symplectic two form
\begin{equation}\label {EC358}
\Omega=dp_r\wedge dr+{\delta}_{ij}dp_i\wedge dx_j+ {\theta f(r)\over
2}\epsilon_{ij}d\dot x_i\wedge d\dot x_j+ {\theta f^{'}(r)\over
4}\epsilon_{ij}\dot x_jd\dot x_i\wedge dr.
\end{equation}
Now reducing the theory to first order by applying the equations of
motion we obtain, to lower order approximation in $\theta$, the
following
\begin{equation}\label {EC363}
p_i=m\dot x_i-{\theta kf(r)\over m}{\epsilon}_{ij} x_j+
{\theta f^{'}(r)p_r\over 2m_r}{\epsilon}_{ij}\dot x_j+{\cal O}(\theta ^2).
\end{equation}
In consequence, the Hamiltonian takes the form
\begin{equation}\label {EC364}
H={p_r^2\over 2m_r}+V(r)+{\rho_i^2\over 2m}+{k\over 2}x_i^2+
{\theta kf(r)\over 2m^2}{\epsilon}_{ij}x_i\rho_j+{\cal O}(\theta ^2).
\end{equation}
Using these approximations in the symplectic two-form,
we obtain the following algebra
\[
\{r,p_r\}_D=1,\hspace{0.3cm}\{x_i,\rho_j\}_D={{\delta}_{ij}}+
{\theta f^{'}(r)p_r\over mm_r}{\epsilon}_{ij}
\]
\[
\{p_r,\rho_i\}_D={\theta f^{"}(r)p_r\over mm_r}{\epsilon}_{ij}
\rho_j-{2\theta kf^{'}(r)\over m}{\epsilon}_{ij}x_j,
\]
\[
\{x_i,x_j\}_D={\theta f(r)\over m^2}{\epsilon}_{ij},
\hspace{0.3cm}\{p_r,x_i\}_D={\theta f^{'}(r)\over
2m^2}{\epsilon}_{ij}\rho_j,
\]
\begin{equation}
\{\rho_i,r\}_D={\theta f^{'}(r)\over
mm_r}{\epsilon}_{ij}\rho_j,\hspace{0.3cm}\{\rho_i,\rho_j\}_D={2\theta
kf(r)\over m}{\epsilon}_{ij}.
\end{equation}
where we have used the redefinition $\rho_i=m\dot x_i$. To quantize
this system it is possible to follow similar steps to the performed
in Sec. 2.1, also can be useful to use some of the ideas reported in
\cite{Kupri}, since the noncommutativity parameter is not constant.
\section{Conclusions}
We have derived the relationship between the higher order theories
and noncommutativity using a perturbative method. The relationship
was shown for a Quantum Mechanics generalization of Chern-Simons,
was proved that the noncommutativity arises naturally when we
project the states to the low energy states of the high order
theory. It is interesting to compare the results of this paper and
Ref. \cite{Jian}, who studied the same model, a Hamiltonian similar
to (\ref{EC310}) and commutation relations related to (\ref{EC37}).
However, they proposed the model directly, whereas in our case this
model is the result of a high order time derivative theory.
Furthermore, in our model, unlike its counterpart of first order, it
is not necessary to cancel the kinetic term and the noncommutativity
arises automatically from the projection to lower energy states.
Thus in this example we have shown that the noncommutativity can be
seen as a result of make sense, by perturbation theory, of a high
order time derivative theory. Also, we have derived that this result
is extended to the case of high order time derivative theories and
in the case of a non constant noncommutative parameter.

\section*{Acknowledgments}
The authors acknowledge partial support from DGAPA-UNAM grant PAPIIT
-IN111210 and PROMEP/103.5/13/9043 (O.S.).

\end{document}